\documentstyle[11pt,newpasp,twoside,epsf]{article}
\def\edcomment#1{\iffalse\marginpar{\raggedright\sl#1\/}\else\relax\fi}
\reversemarginpar

\begin{document}
\title{Results from Microlensing Searches for Extrasolar Planets}
 \author{Penny D. Sackett}
\affil{Kapteyn Institute, 9700 AV Groningen, The Netherlands}

\begin{abstract}
Specially-designed microlensing searches, some of which have been underway 
for several years, are sensitive to extrasolar planets orbiting the most 
common stars in our Galaxy. Microlensing is particularly well-suited 
to the detection of Jupiter-mass planets orbiting their parent stars at several AU. 
Since Jovian analogs are thought to influence the subsequent evolution of most 
planetary systems, they are particularly important to study. 
The orbital radii and distances to the planetary systems probed 
by microlensing are larger than those currently studied by 
radial velocity techniques; the two methods are thus complementary.
Recent results from microlensing searches are discussed, 
including constraints on Jovian analogs orbiting typical Galactic 
stars.  Benefits and drawbacks of the technique for the 
characterization of planetary systems, and future prospects  
are briefly reviewed. 
\end{abstract}

\section{Introduction}

The first detection of an extrasolar planet 
around a normal star (Mayor \& Queloz 1995) and    
subsequent deluge of similar discoveries by the radial velocity 
technique (Marcy, Cochran \& Mayor 2000), 
have taught us that $\sim$5\% of solar type 
stars harbor planetary systems very {\it unlike\/} our own.  
What remains to be determined is the abundance of planetary 
systems {\it similar\/} to that of the Sun's (eg., terrestrial planets 
at $\sim$0.5-2~AU or Jovian analogs at $\sim$3-12~AU) and the frequency  
of planets around the most common stars in our Galaxy, M dwarfs.  
Microlensing is providing the first partial answers 
to these questions. 

\section{Galactic Planetary Systems As Multiple Microlenses}

Microlensing occurs whenever a massive compact object (such as a star, 
black hole, etc.) 
passes very near the line-of-sight to a background luminous source.
In the case of a single point lens, two images of the 
source are formed with a separation that scales with the 
angular radius of the Einstein ring, 
$\theta_{\rm E} \equiv [4 G M (1 - x)/(c^2 D_L)]^{-1/2}$, 
where $x \equiv D_L/D_S$, and $M$, $D_L$, and $D_S$ are the mass of 
the lens, its distance, and the distance to the source, respectively. 
For sources in the 
Galactic bulge, this separation is $\sim$1~mas,  
and thus generally too close to be resolved.  The combined image 
brightness, however, can be observed and is a function of the changing 
projected distance between the lens and the observer-source line-of-sight.
As the source moves through the axisymmetric magnification pattern 
generated by a single lens, 
a symmetric light curve results.

This symmetry is destroyed for multiple lenses such as binary star 
systems or planetary systems.  More complicated magnification patterns 
are formed, and the resulting light curve depends on the angle $\phi$ of 
the source trajectory. For binary lenses, the 
topology of the magnification map (Fig.~1) depends on only two parameters: 
the mass ratio $q \equiv m_1/m_2$ of the lenses and their 
projected separation $d \equiv a_p/R_{\rm E} = a_p/(D_L \, \theta_{\rm E})$ 
in units of the Einstein ring radius.  For typical Galactic lenses,  
M dwarfs in the bulge or inner disk, 
the physical size of the Einstein radius is $R_{\rm E} \sim 2-3$~AU. 
Because typical $R_E$ are comparable to the 
orbital radius of many Solar System planets, microlensing 
is an excellent way to search for planets, as first suggested by 
Mao \& Paczyn\'ski (1991) and further developed by Gould \& Loeb (1992).

\begin{figure}
\hglue 0.0cm\epsfxsize=12.5cm\epsffile{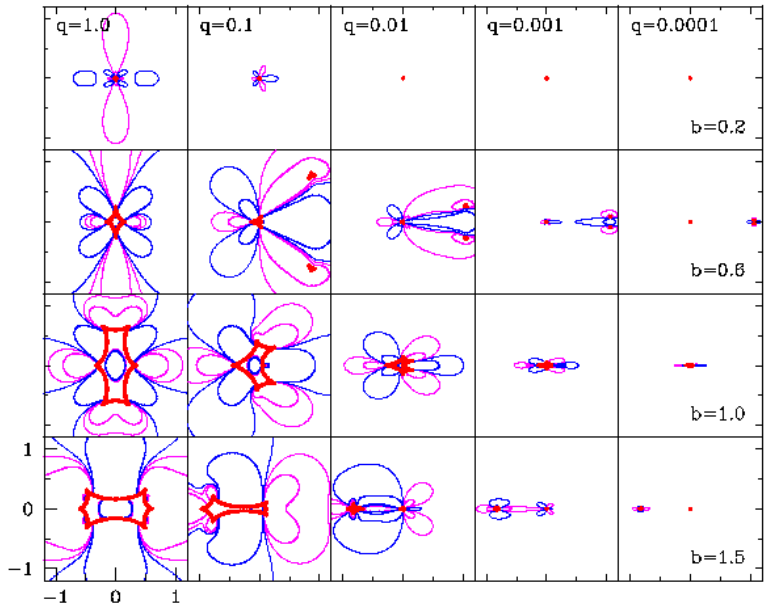}
\vglue -4.25cm
\caption{Magnification patterns for binary lenses with 
mass ratio $q$ and projected separation $d$ ($=b$ in this figure). 
{\it Differences\/} in magnification of $\pm1\%$ and $\pm5\%$ 
between a point lens and a binary of the same total mass are 
shown as contours.  The heaviest contours are the ``caustic 
curves.''  A Jupiter-Sun lensing system would have $q \approx 10^{-3}$ and 
$d \approx 1.2$ if located halfway between the observer and the 
Galactic bulge and viewed face-on.  
(Adapted from Gaudi \& Sackett 2000).
}
\vglue -0.5cm
\end{figure}

Figure~1 shows the differential magnification effect of binary lenses 
as a function of their mass ratio $q$ and separation $d$.  
Each panel covers an area $\sim \theta_{\rm E}$ on a side 
centered on the more massive primary lens.  
The brightness of an image is proportional to 
$| det J |^{-1}$, where $J$ is the Jacobian that describes 
the coordinate transformation between the image and source planes.  
The locus of points in the source plane for which $| det J | = 0$ 
is called the ``caustic curve.''  A source crossing the caustic 
will be highly magnified. 

\begin{figure}
\vglue -8.5cm
\hglue -0.5cm\epsfxsize=14cm\epsffile{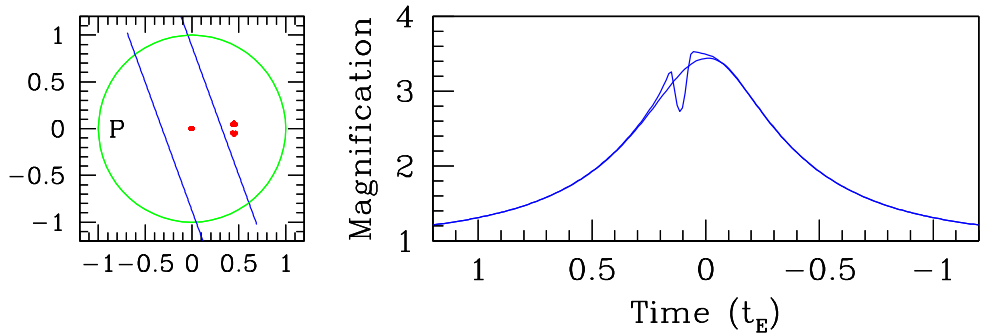}
\vglue -1cm
\caption{Left: The position of a $q=10^{-3}$ planet is marked with ``P''  
inside the Einstein ring radius of its primary at $d=0.8$.  The central 
caustic and two ``planetary caustics'' are shown, 
along with two possible source trajectories.  Note that the 
caustics do {\it not\/} coincide with the position of planet.  
Right: The resulting light curves. 
}
\end{figure}

Since most microlensing events are alerted in real time only 
when the source lies near or inside $\theta_{\rm E}$, 
a light curve will reveal the binarity of the lens 
(via an anomaly $> 1\%$) only if the source trajectory passes over 
the contours shown in Fig~1.  Otherwise, the light curve will be 
indistinguishable from that due to a single isolated lens.  
Figure~2 illustrates this effect; two possible source 
trajectories are shown passing the same distance from the ``central''
caustic, but different distances from the two, outer ``planetary'' 
caustics (all three caustics are caused by the lens-planet combination).
The light curve associated with only one of these trajectories 
betrays the presence of the planet.  Since the deviation contours 
are most extended for $q$ and $d$ of order unity (Fig.~1), these 
types of binary lenses will be the easiest to detect.   

\subsection{Sensitivities vs Efficiencies}

Even planets as small as the Earth generate caustic structures as 
they orbit Galactic lenses; background sources passing over or near 
these caustics will experience a substantial additional microlensing 
effect due to the binarity of the lens.  
The cross section of the planetary caustic and thus the probability 
that a random trajectory will pass over or near it decreases slowly 
with mass ratio (Dominik 1999).  
For this reason, microlensing --- unlike most other planet detection 
methods --- has {\it sensitivity\/} to terrestrial-mass planets.  
This sensitivity is limited primarily by the sampling rate and precision of 
the photometry required to see signals that 
are reduced in amplitude by finite source effects for $q \le 10^{-4}$. 
The {\it efficiency\/} with which planets of given $q$ and $d$ can be detected 
in a given light curve is a statistical quantity that depends 
on the photometric precision, sampling and duration of the 
photometric data and on the characteristics of the underlying event 
sampled, most notably its minimum impact parameter, $u_o$, and 
the amount of blended non-lensed light in the same resolution 
element (Gaudi \& Sackett 2000).   Determination 
of detection efficiencies for individual light curves 
allows non-detections to be translated into statistical upper limits 
for a given class of planet.

\subsection{Detection vs Characterization}

In principle, the {\it detection\/} of a microlensing anomaly can 
be quantified statistically, 
but its clear identification with a planetary lens relies on the 
{\it characterization\/} of the planet through determination 
of $q$ and $d$.  Since any given event will not repeat, characterization 
must be obtained at the time of detection, and will 
require better data.  
With sufficient data quality, both $d$ and $q$ 
can be determined from light curve modeling, although in some  
cases the well-known $d - d^{-1}$ ``degeneracy'' (Griest \& 
Safizadeh 1998; Dominik 1999) will prevent a unique identification.  
High magnification events 
are sensitive and reasonably efficient to the detection of small mass 
planets since their small $u_o$ brings them close to the central 
caustic generated by companions (Griest \& Safizadeh 1998).  
Unfortunately, such detections 
may be particularly difficult to characterize, since all planets orbiting 
the primary lens will distort the central caustic 
(Gaudi, Naber \& Sackett 1998); more massive ones 
will have an effect over a larger range of $d$.
Di~Stefano and Scalzo (1999) point 
out that current programs typically halt intensive monitoring 
when $u \ge 1$; extended high-precision monitoring may detect 
and characterize planets at larger separation from their host lenses.

\begin{figure}
\hglue 0.25cm\epsfxsize=5cm\epsffile{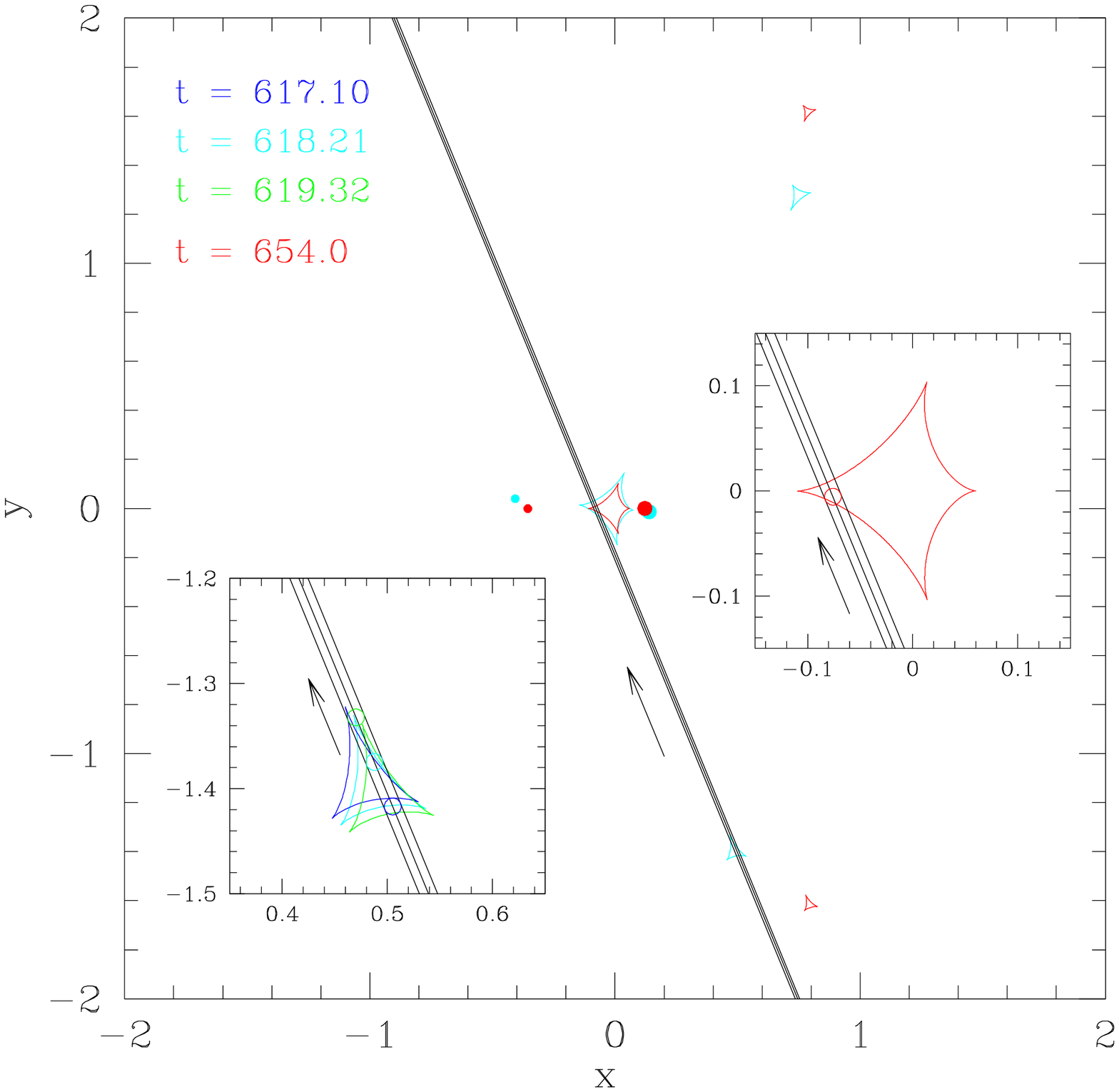}
\vglue -5.30cm
\hglue 5.25cm\epsfxsize=5.5cm\epsffile{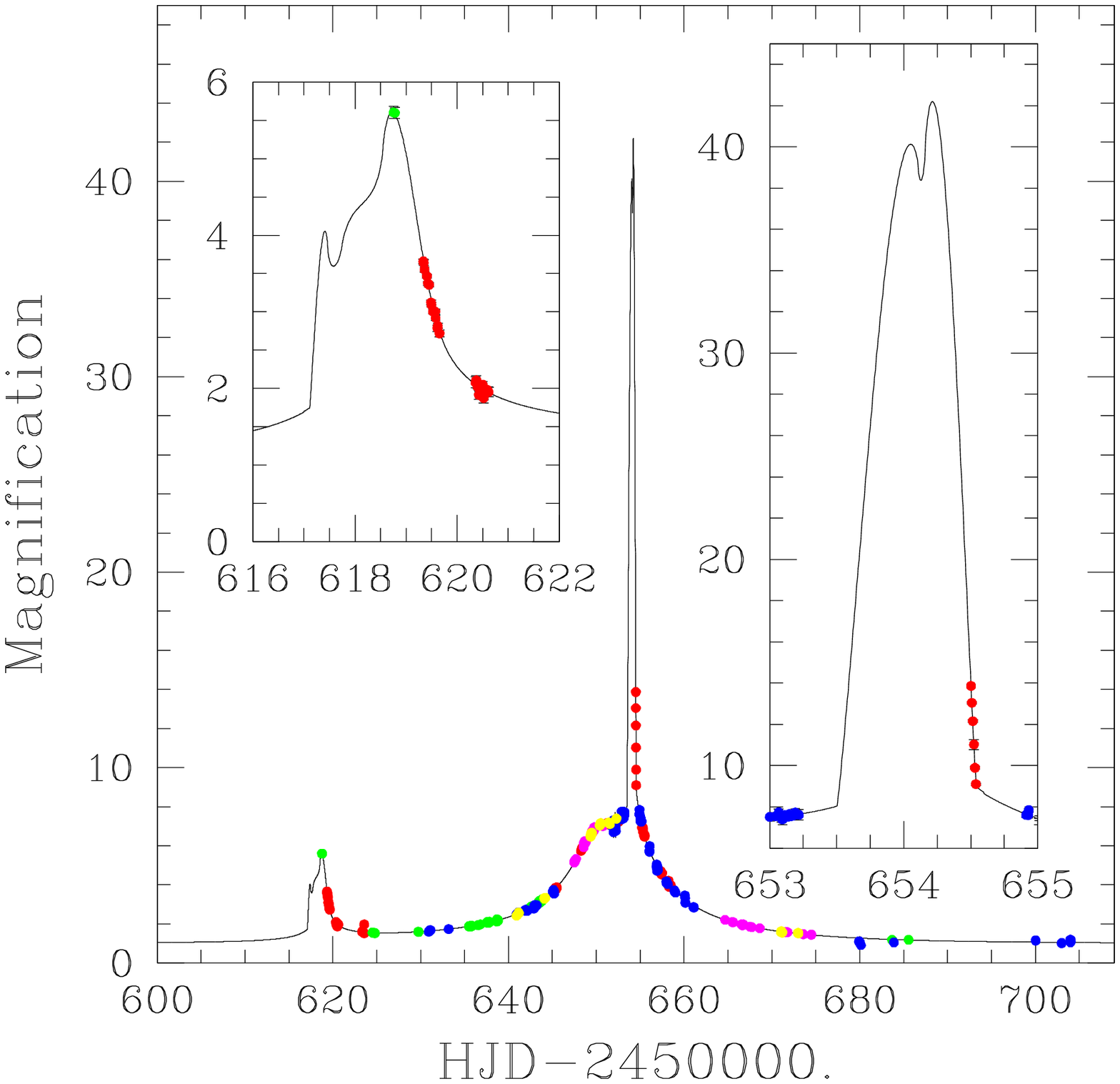}
\vglue -7.25cm
\hglue 8.9cm\epsfxsize=8.5cm\epsffile{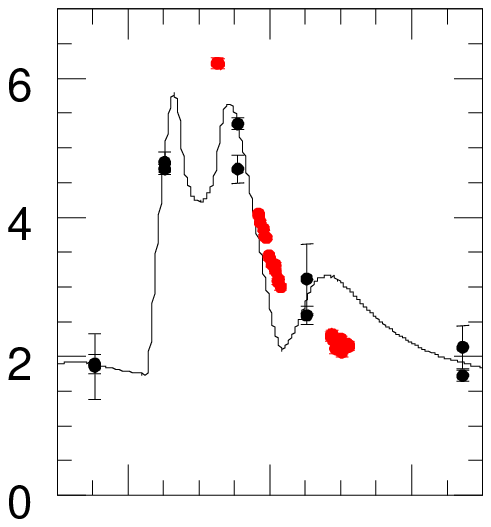}
\vglue -1.75cm
\caption{Left: Moving caustic structure of rotating binary model 
of MACHO 1997-BLG-41. Inserts show zooms; arrow indicates source motion. 
Middle: Rotating binary fit to multi-site PLANET team light curve.
Right: Static binary+planet model atop PLANET (red) and MACHO-GMAN (black) 
data over first caustic feature (\S\ 3.1).
}
\vglue -0.2cm
\end{figure}

\section{Have Exoplanets been Discovered with Microlensing?}

Of 21 light curves consistent with binary lenses analyzed by 
MACHO, two produced acceptable fits 
with companion masses as small $0.05M_\odot$ 
(Alcock et al.\ 2000).  
The first suggestion of possible planetary-mass lenses   
(Bennett et al.\ 1997)   
came from MACHO survey data for MACHO~1994-BLG-4, which could 
be modelled as an M-dwarf/5$M_J$ pair, 
and MACHO~1995-BLG-3, a very short duration event modelled 
as an isolated 2$M_J$ lens.  However, many alternate interpretations  
were allowed by the infrequently sampled light curves inherent 
to survey data (which must sample $\sim$$10^6$ stars nightly!); 
no firm planetary detection was claimed in either case.

The timescale $t_{\rm E} = \theta_{\rm E}/ \mu$ 
($\mu$ is the relative proper motion of lens and source) 
of typical Galactic microlensing events is weeks to months; whereas     
planetary ($q \le 0.01$) anomalies would have 
durations of hours to days.  The necessity of  
round-the-clock monitoring for detection and 
characterization of short-lived planetary deviations prompted the 
establishment of international collaborations such as MPS 
(Rhie et al.\ 2000) and PLANET (Albrow et al.\ 1998),  
which obtain sub-day to sub-hourly photometry on events 
discovered by the survey teams MACHO, OGLE and EROS. Although tantalizing 
hints have been seen in some light curves (see below), 
no clear {\it planetary\/} anomaly has been detected.

\begin{figure}
\hglue 2.25cm\epsfxsize=9cm\epsffile{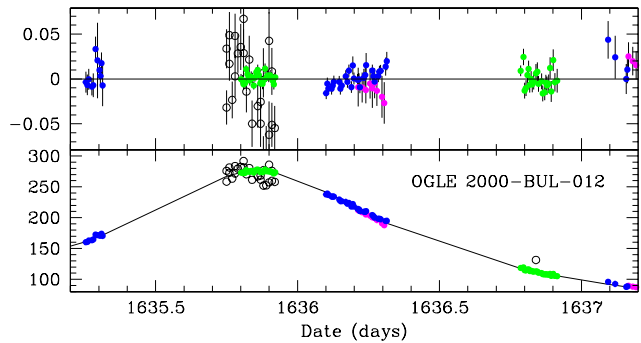}
\vglue -4.0cm
\caption{Bottom: Single-lens fit to OGLE 2000-BUL-12 light curve 
data.  Flux units are linear and arbitrary. Black, open 
circles are OGLE data; colored, filled dots are multi-site PLANET data.  
Top: Residuals from the single-lens model 
expressed as a fraction of the flux.  
}
\vglue -0.25cm
\end{figure}

{\bf MACHO 1997-BLG-41: }
The very unusual event MACHO 1997-BLG-41 caused a stir in the 
community: although clearly multiple lens microlensing,  
static stellar binary models did not fit the data.  Bennett et al.\ (1999) 
interpreted the double caustic structure as coming from 
a static triple system, a Jovian planet orbiting a stellar binary. 
The PLANET team, modeling their own data, instead showed that the 
light curve could be fit as a normal stellar binary (Albrow et al.\ 2000a) 
whose rotation brings one of the triangular caustics across the 
source trajectory (Fig.~3).  The proposed binary+Jovian model 
was incompatible with PLANET data.

{\bf OGLE 2000-BUL-12: }
Recently, Yock et al.\ (2000) have suggested that a planet 
may be orbiting the primary lens of event OGLE~2000-BUL-12, based 
on analysis of public domain OGLE data (Udalski \& Szymanski 2000) 
that displayed a possible anomaly at peak.  Unpublished PLANET team 
data indicates no anomaly and much smaller scatter 
at the same temporal position (Fig.~4).

\begin{figure}
\hglue 0.5cm\epsfxsize=6.5cm\epsffile{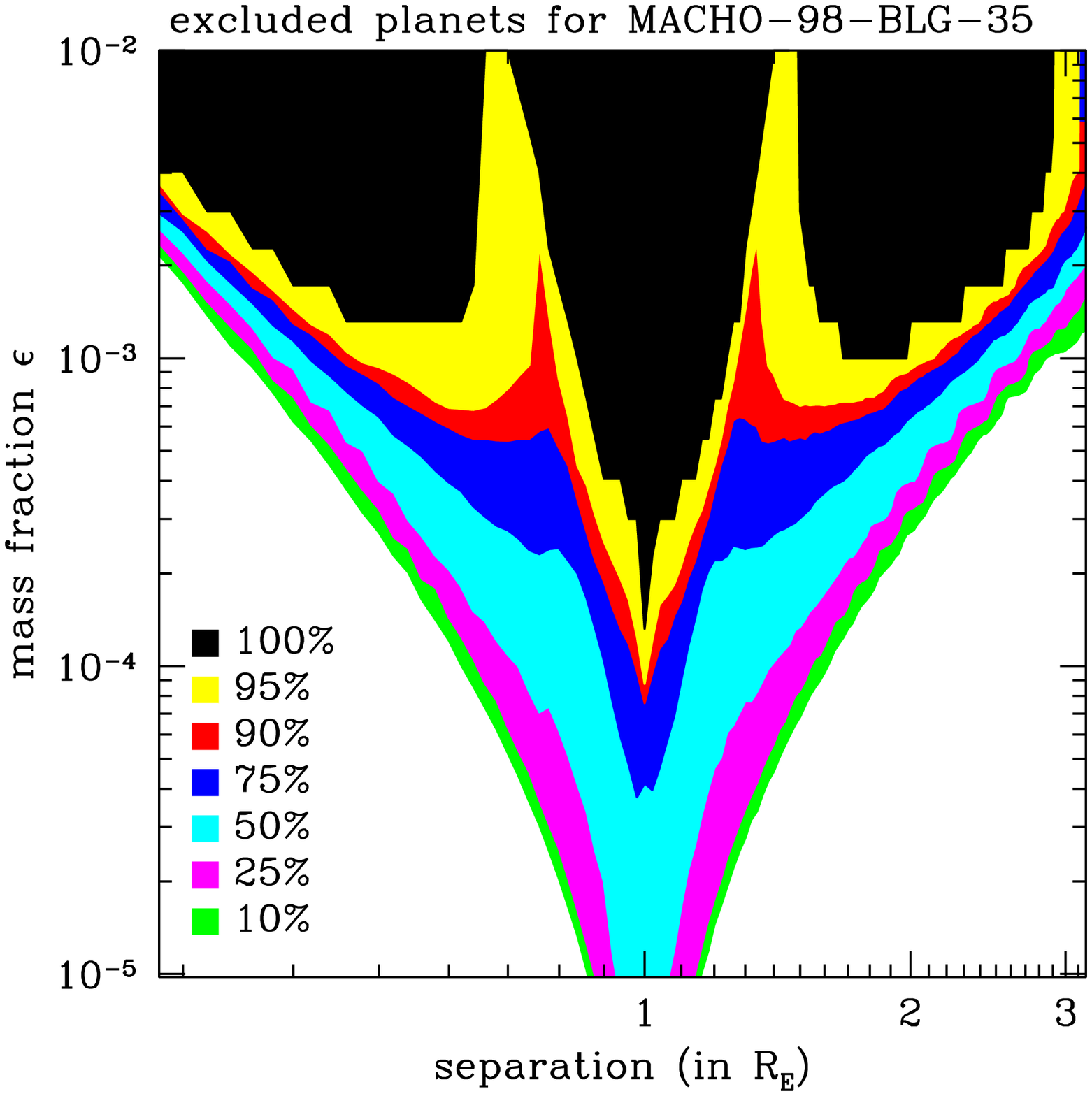}
\vglue -6.75cm
\hglue 6.75cm\epsfxsize=6.75cm\epsffile{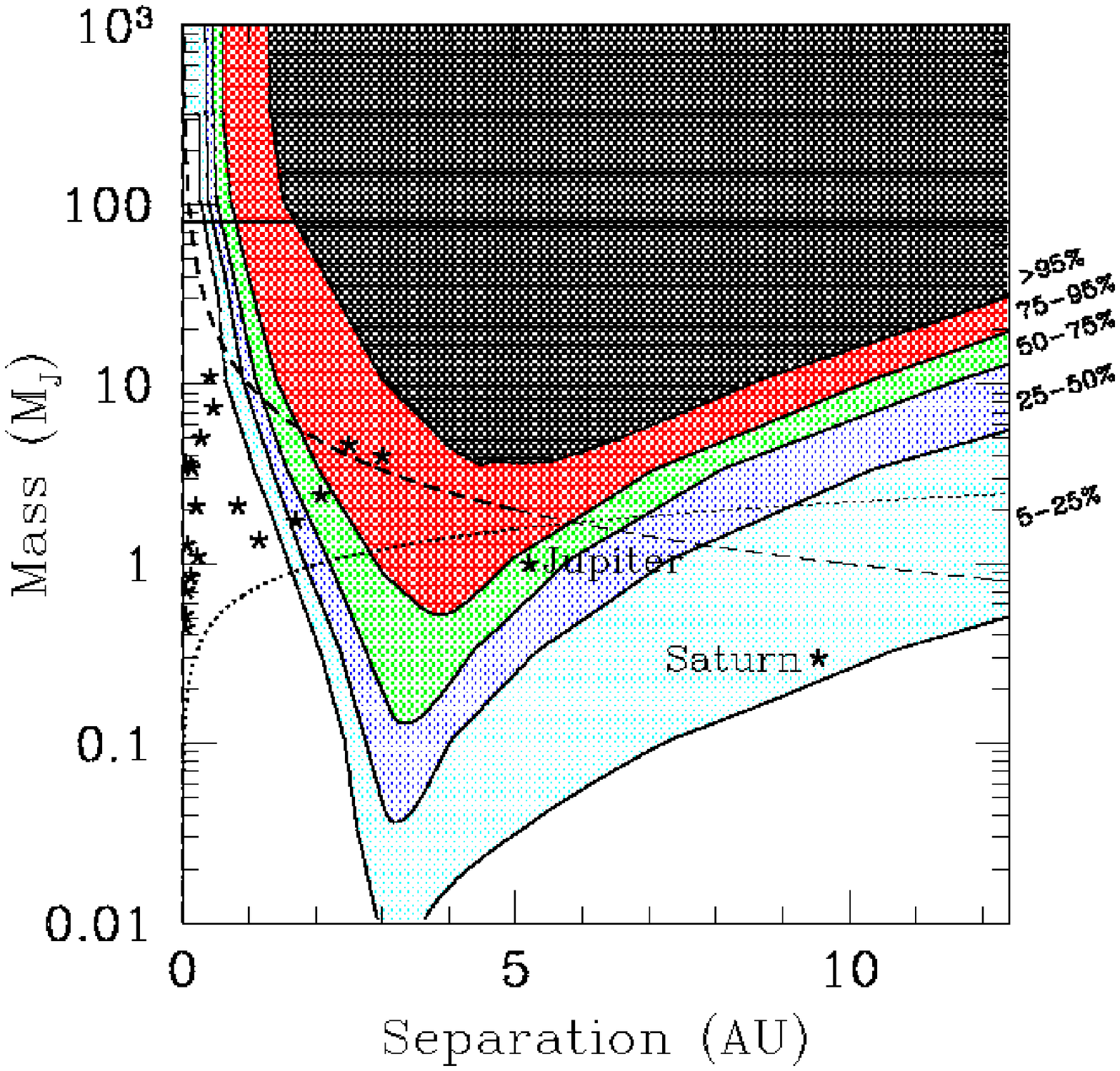}
\vglue -0.25cm
\caption{Left: Planetary exclusion contours at various significance 
levels as a function of mass fraction and $d$ for companions 
to event MACHO~98-BLG-35 (Rhie et al.\ 2000).  Right: 
Same for event OGLE~1998-BUL-14, where the planetary mass 
and orbital separation have been estimated assuming 
a 1M$_\odot$ mass lens with $R_{\rm E} = 3$AU (Albrow et al.\ 2000b).
Asterisks mark known radial velocity planets.}
\vglue -0.25cm
\end{figure}

{\bf MACHO 1998-BLG-35: }
Although ruling out a large class of high-mass planets orbiting 
the lens of MACHO~1998-BLG-35 (\S4.1), the MPS 
and MOA groups (Rhie et al.\ 2000) noticed a weak anomaly near 
the peak of this 
high amplification event that they interpreted as intriguing 
evidence of a low-mass companion with 
$4 \times 10^{-5} < q < 2 \times 10^{-4}$.  
The signal fell below the formal MPS/MOA detection limit.  
PLANET team data for this event did not confirm this detection; 
the PLANET light curve was consistent with an 
isolated point lens (Albrow et al.\ 2000c).

\section{Limits on the Abundance of Jovian Planets in the Galaxy}

Current microlensing planet searches have appreciable efficiency 
for the detection of companions with mass ratio $q \ge 10^{-4}$, 
ie., planets with masses of order $m_p \ge 0.1M_J$.  The lack of 
detected perturbations consistent with planets in this mass range allows 
constraints to be placed on the abundance of Jovian analogs  
around typical stars (ie, lenses) in the Milky Way.

The first published exclusion diagram (Fig.~5) 
for companions in an individual lensing 
system was presented by the MPS/MOA collaborations for MACHO~98-BLG-35 
(Rhie et al.\ 2000).  Due to the high amplification of this event, 
the source would have passed very near any (central) 
caustic structure due to massive companions.  Since no anomaly 
consistent with Jovian-mass companions was seen, exclusion contours 
could be derived as a function of $q$ and $d$ for this 
lensing system (Fig.~5).  The PLANET collaboration performed 
a similar analysis for another high amplification event, 
OGLE~1998-BUL-14 (Albrow et al.\ 1999b).  When converted to 
parameters appropriate to a solar-type lens in the bulge and 
averaged over all orbital inclinations, the exclusion contours for 
Jovian and super-Jovian planets in the OGLE~1998-BUL-14 
overlap the region of parameter space inhabited by 
current planet detections by the radial velocity technique (Fig.\ 5). 
For the most part, however, microlensing is sensitive planets at 
larger orbital radii (1-10~AU) than is the Doppler technique.
 
\begin{figure}
\hglue 0.5cm\epsfxsize=6.5cm\epsffile{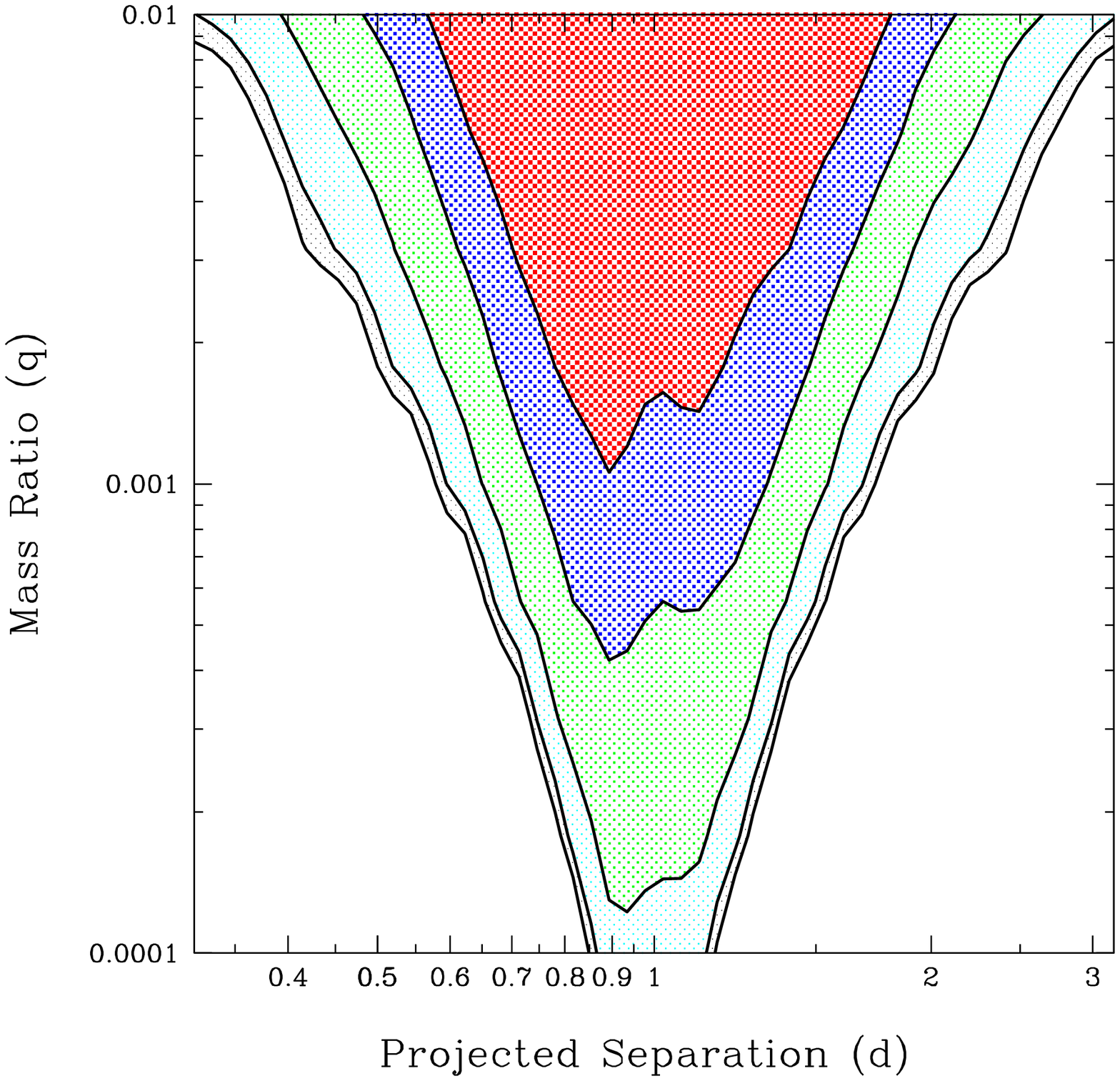}
\vglue -6.5cm
\hglue 6.5cm\epsfxsize=6.5cm\epsffile{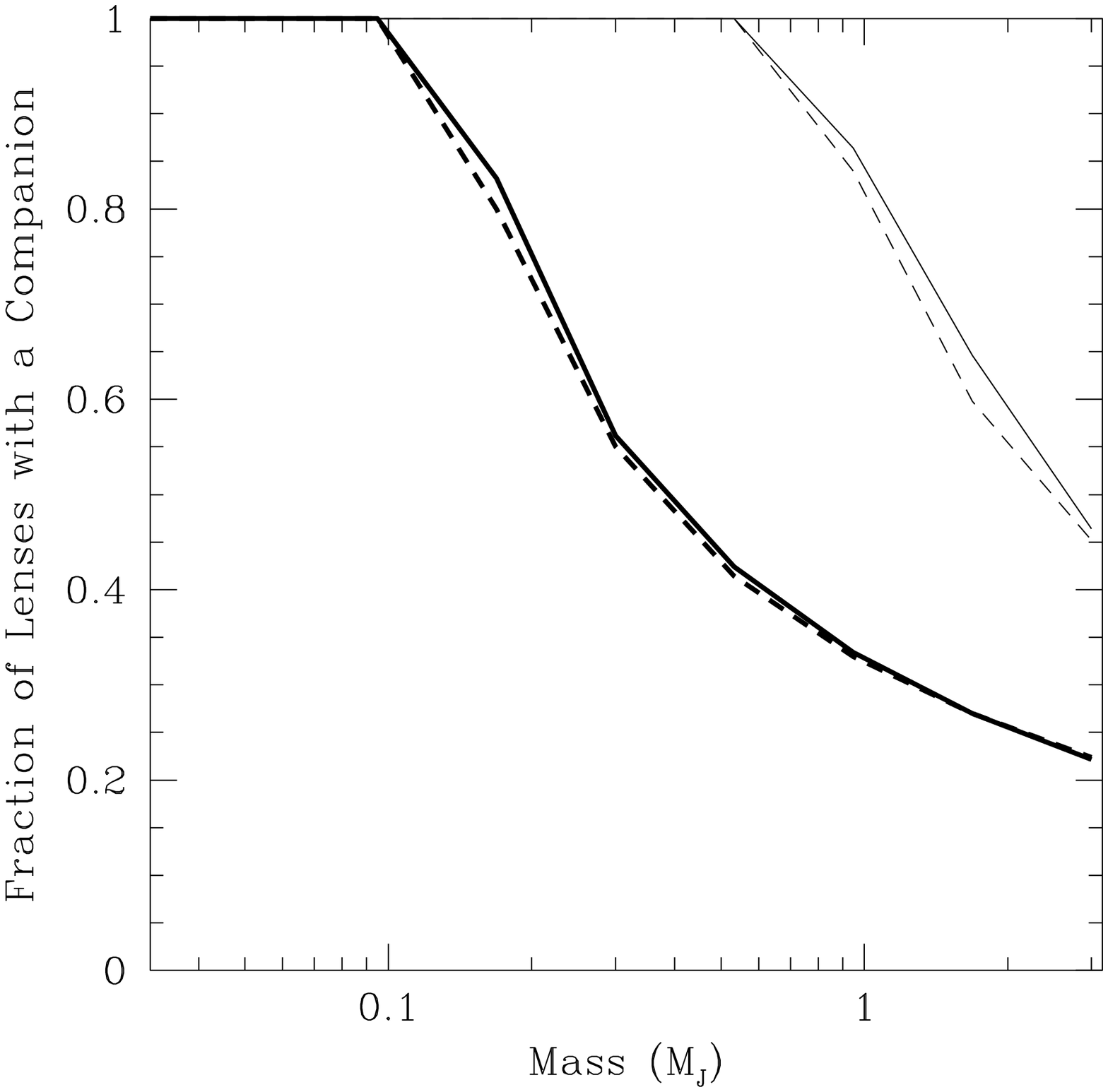}
\vglue -0.25cm
\caption{Left: Upper limits (95\% significance) on the presence 
of companions in a given fraction $f$ of microlenses, shown as contours 
in mass ratio ($q$) -- projected separation ($d$) space.  
Statistically, $q=0.001 \rightarrow m_P \approx 0.3$M$_{\rm Jup}$ and 
$d = 1 \rightarrow r_P \approx 2$AU. Right: Upper limits on $f$ as a 
function of companion mass for orbital radii in the indicated 
ranges (adapted from Albrow et al.\ 2000c).   
The negligible effect of finite source size on the 
efficiency calculations is also shown (dashed). 
}
\vglue -6.6cm
{\small \hglue 5.1cm $f = 1/4$
\vglue 0.05cm\hglue 5.4cm  $= 1/3$
\vglue 0.05cm\hglue 5.4cm  $= 1/2$
\vglue 0.05cm\hglue 5.4cm  $= 2/3$
\vglue 0.05cm\hglue 5.4cm  $= 3/4$
\vglue -0.75cm
\hglue 9cm 1.5 AU$ \, < a < \, $4 AU
\vglue -2.5cm\hglue 10.2cm  1$\, < a < \, $7 } 
\vglue 6.75cm
\end{figure}

Based on 43 microlensing events collected from five years of photometric 
observations, the PLANET team has recently announced constraints on 
the abundance of Jovian and super-Jovian planets orbiting typical stars 
(ie., lenses) in the Galaxy (Gaudi 2000, Albrow et al.\ 2000c).  
No clear planetary anomalies were observed in these 43 events, 
implying that less than one-third of $\sim$0.3~M$_\odot$ stars have 
Jovian-mass companions with semi-major axes in the range 
1.5 AU$\, < a < \, $4 AU (Fig.\ 6).  Weaker limits are placed 
on the existence of Jovian planets orbiting in the range 
1 AU$\, < a < \, $7 AU.  These are the first constraints on  
exoplanets orbiting the most common of Galactic stars: M-dwarfs. 

\section{Prospects for the Future}

As with all techniques, microlensing has advantages and disadvantages.
In the cadre of exoplanet search techniques, it offers the opportunity 
to {\it detect and characterize Jovian planets at large orbital radius 
(1-10~AU) without waiting for one or more orbital periods\/}.  
Currently, it also has more {\it sensitivity to Neptune-mass 
planets\/} than any other ground-based search technique.  In the future, 
this sensitivity may be enhanced and extended with new facilities 
on the ground (Peale 1997, Sackett 1997) and in space (Rhie, this conference).  
Several {\it challenges must be overcome to detect and characterize
Earth-mass planets with reasonable efficiency\/}, including 
the diluting effect of finite source size on the magnification 
gradient due to  
small caustics (Bennett \& Rhie 1996, Gaudi \& Sackett 2000).

Currently, the mass ratio (but not the mass) and the projected 
separation in units of $\theta_{\rm E}$ (but not in AU) of a lensing planet 
can be determined from an anomaly.  
More information may be forthcoming in the future.  
{\it Large-aperture spectroscopy\/}  
of an event may yield 
spectroscopic identification of the faint primary lens against the sea 
of source star light (Mao, Reetz \& Lennon 1998), 
thereby allowing the 
mass of the primary and $\theta_{\rm E}$ to be estimated. 
Measurement of the {\it photometric centroid motion\/} 
due to the changing position and brightness 
of the microimages (astrometric microlensing), could yield 
the mass of the lens (Dominik \& Sahu 2000, Han 2000, Gould \& Han 2000), 
and a more robust determination of planetary parameters 
(Safizadeh, Dalal \& Griest 1999). 

\acknowledgments

I am grateful to my PLANET colleagues for permission to display unpublished 
data for OGLE 2000-BUL-12.


\begin{references}

\reference Albrow, M.D., et al.\ 2000a, \apj, 534, 894

\reference Albrow, M.D., et al.\ 2000b, \apj, 535, 176

\reference Albrow, M.D., et al.\ 2000c, \apj\ Letters, submitted 
(astro-ph/0008078) 

\reference Alcock, C. et al.\ 2000, \apj, 541, 270

\reference Bennett, D.P. \& Rhie, S.H. 1996, 472, 660

\reference Bennett, D.P., et al.\ 1997, ASP Conf.\ Prof.\ 119, 95 (astro-ph/9612208)

\reference Bennett, D.P., et al.\ 1999, Nature, 402, 57

\reference Di~Stefano, R., \& Scalzo 1999, \apj, 512, 579

\reference Dominik, M., 1999, \aap, 349, 108

\reference Dominik, M., \& Sahu, K. 2000, \apj, 534, 213

\reference Gaudi, B.S. 2000, PhD Thesis, The Ohio State University

\reference Gaudi, B.S., Naber \& Sackett, 1998, \apj\ Letters, 502L, 33

\reference Gaudi, B.S., \& Sackett, P.D. 2000, \apj, 528, 56

\reference Gould, A., \& Han, C. 2000, \apj, 538, 653

\reference Gould, A., \& Loeb, A. 1992, \apj, 396, 104

\reference Griest, K., \& Safizadeh, N. 1998, \apj, 500, 37

\reference Han, C., in A New Era of Microlensing Astrophysics, 
	ASP Conf. Series, 
	eds. Menzies \& Sackett, in press (astro-ph/0003369)

\reference Marcy, G., Cochran, W.D. \& Mayor, M. 2000, 
	in Protostars and Planets IV, eds. Mannings, Boss \& Russell 
(U. of Arizona Press: Tucson) p. 1285

\reference Mayor, M. \& Queloz, D. 1995, Nature, 378, 355

\reference Mao, S. \& Paczyn\'ski, B. 1991, \apj\ Letters, 374L, 37

\reference Mao, S., Reetz, J., \& Lennon, D.J. 1998, \aap, 338, 56 

\reference Peale, S. 1997, Icarus, 127, 269

\reference Rhie, S.H. et al.\ 2000, \apj, 533, 378

\reference Sackett, P.D. 1997, ESO Document: SPG-VLTI-97/002 (astro-ph/9709269)

\reference Safizadeh, N., Dalal, N. \& Griest, K. 1999, \apj, 522, 512

\reference Udalski, A. \& Szymanski, M. 2000, 
	http://www.astrouw.edu.pl/$~$ftp/ogle/ogle2/ews/ews.html

\end{references}
\end{document}